\documentclass[10pt,a4paper,conference]{IEEEtran}

\usepackage{graphicx}
\usepackage{bm}
\usepackage{cite}
\usepackage{amsmath}
\usepackage{amsfonts}
\usepackage{amssymb}
\usepackage{color}
\usepackage[a4paper]{geometry}



\title{\LARGE{Atomistic modeling of the \\ phonon dispersion and lattice properties of \\ free-standing  $\langle$100$\rangle$ Si nanowires}}%

\author{\authorblockN{%
	Abhijeet Paul, Mathieu Luisier and Gerhard Klimeck}
	\authorblockA{%
	School of Electrical and Computer Engineering and Network for Computational Nanotechnology, \\
	Purdue University, West Lafayette, IN, USA- 47906, e-mail: paul1@purdue.edu}
}%

\begin{document}

\maketitle 
\vspace{-0.6cm}

\begin{abstract}
 Phonon dispersions in $\langle$100$\rangle$ silicon
 nanowires (SiNW) are modeled using a
 Modified Valence Force Field (MVFF) method based on atomistic force constants. 
 The model replicates the bulk Si phonon dispersion very well. In SiNWs, apart from four acoustic
 like branches, a lot of flat branches appear indicating strong phonon confinement in these nanowires
 and strongly affecting their lattice properties. The sound velocity ($V_{snd}$)
 and the lattice thermal conductance ($\kappa_{l}$) decrease as the wire cross-section size is reduced 
 whereas the specific heat ($C_{v}$) increases due to increased phonon confinement and surface-to-volume ratio (SVR).
\end{abstract}

\section{Introduction}
 \label{par:1}
Silicon nanowires (SiNWs) are playing a vital role in areas ranging from CMOS \cite{ITRS} to thermo-electricity \cite{thermo}. The finite extent and increased surface-to-volume ratio (SVR) in these nanowires result in very different phonon dispersions compared to the bulk materials \cite{anantram_ph,continuum_model}. This work investigates the effect of geometrical confinement on the phonon dispersion, the sound velocity ($V_{snd}$), the specific heat ($C_{v}$), and the lattice thermal conductance ($\kappa_{l}$) in $\langle$100$\rangle$ SiNWs using a Modified Valence Force Field (MVFF) model based on atomistic force constants.

Previous theoretical works have reported the calculation of phonon dispersions in SiNWs using a continuum elastic model and Boltzmann transport equation \cite{continuum_model}, atomistic first principle methods like DFPT (Density Functional Perturbation Theory) \cite{SINW_110_phonon,jauho_method,sinw_cv,strain_effect_1} and atomistic frozen phonon approaches like Keating-VFF (KVFF) \cite{Mingo_kappa,mahan_phonon}.  Thermal conductivity in SiNWs has been studied previously using the KVFF model \cite{Mingo,strain_effect_2}. The continuum approaches fail to capture the atomistic effects like anisotropy in $C_{v}$ for nanowires, etc., whereas the first principle models cannot be extended to very large structures due to their heavy computational requirement. The VFF model overcomes these shortcoming by capturing the atomistic effects and solving phonon dispersions and transport in realistic structures \cite{Mingo_kappa, Mingo}.  However, a simple KVFF model does not reproduce the bulk dispersions very well and hence cannot be used confidently for nanostructures \cite{jce_own_paper}.  The MVFF model overcomes the verifiable shortcomings in bulk. We use the MVFF model for phonon calculations in SiNWs and believe them to be more robust and predictive than the KVFF model \cite{jce_own_paper}. 


\begin{figure}[t!]
	\centering
		\includegraphics[width=3.1in,height=2.3in]{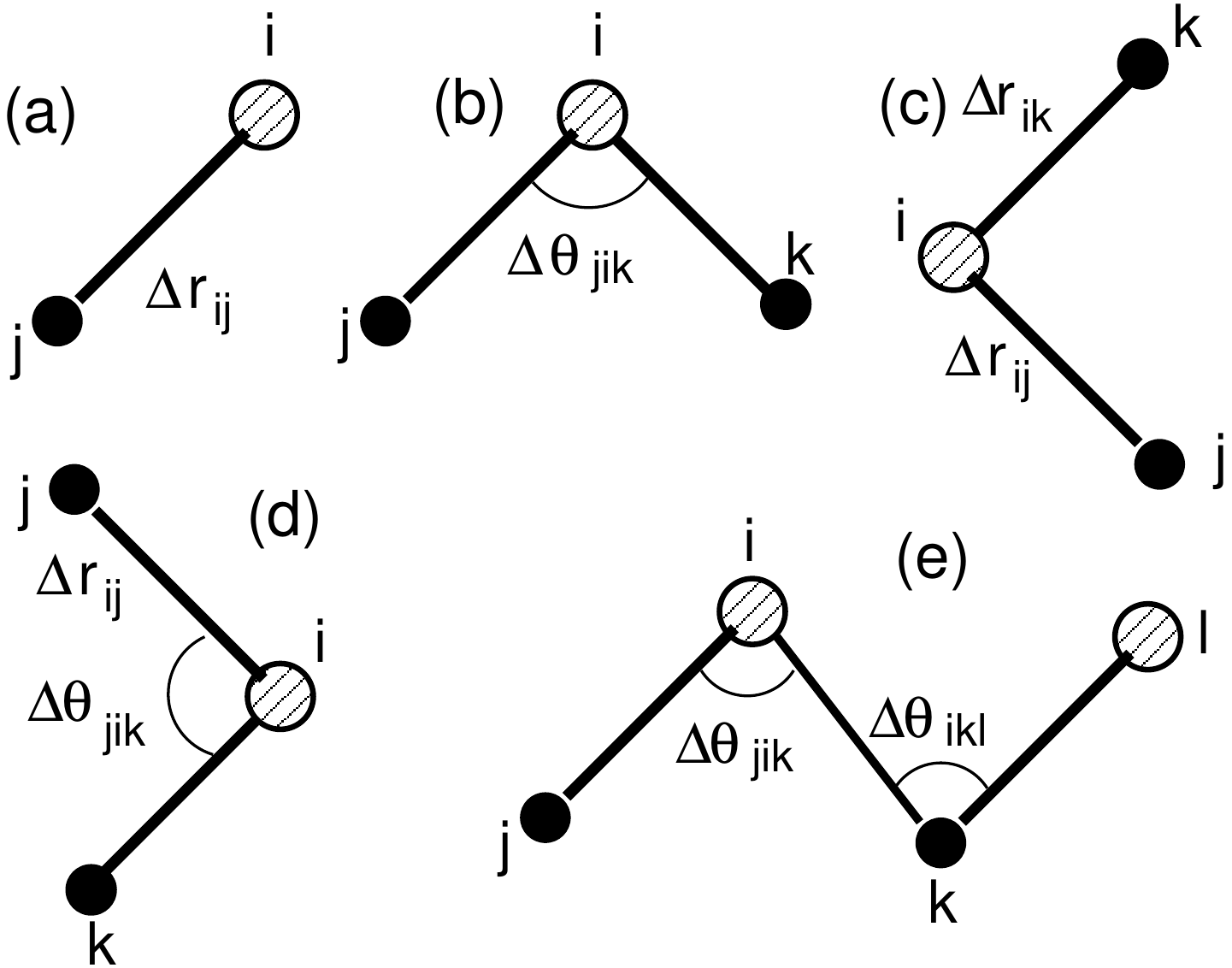}
	\caption{The short range interactions used for Phonon dispersion calculations. The interactions are (a) bond-stretching($\alpha$), (b) bond-bending($\beta$) (c) bond stretch coupling($\gamma$) (d) bond bending-stretching coupling ($\delta$) (e) coplanar bond bending coupling($\lambda$).
	}
	\label{fig:VFF_short_range}
\end{figure}  

\begin{figure}[b!]
	\centering
		\includegraphics[width=1.9in,height=1.9in]{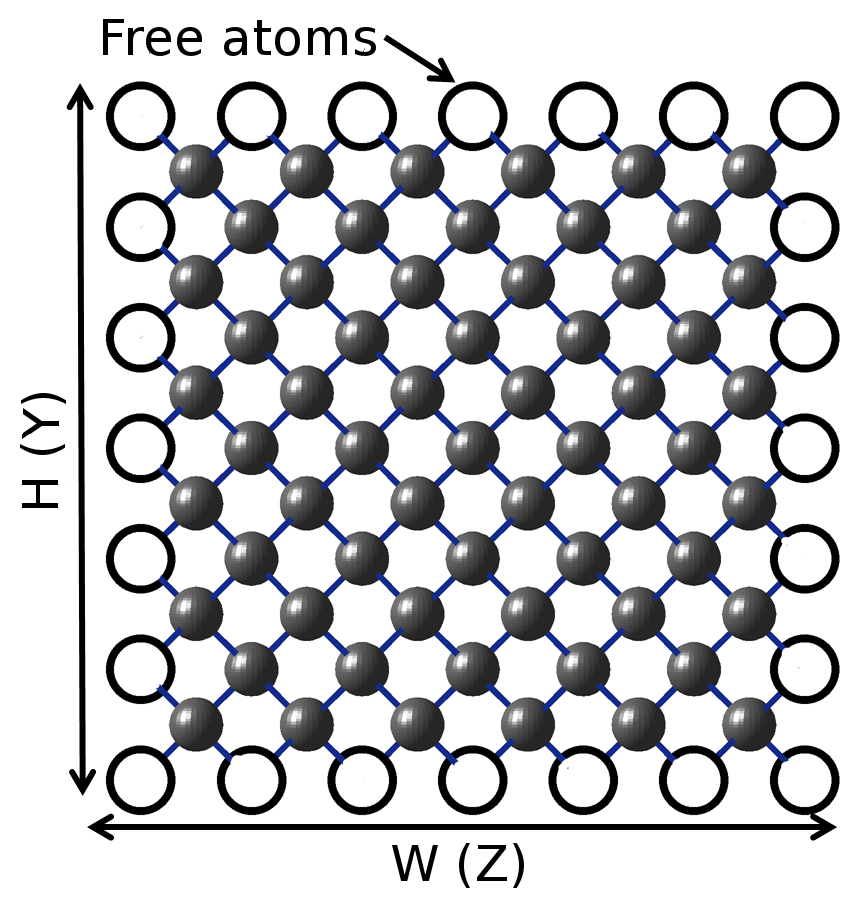}
	\caption{Projected unitcell structure of a $\langle$100$\rangle$ oriented Silicon nanowire. The white atoms (for visual guidance) show the freely vibrating surface atoms. Grey colored atoms are inside the wire}
	\label{fig:Si_NW_100}
\end{figure}
This paper is organized in the following sections. Section \ref{par:2} briefly presents the phonon model and its application to SiNWs. Also the computation of different wire lattice properties is outlined there. The results on phonon dispersion and other lattice properties are presented in Sec. \ref{sec:3}. Conclusions are summarized in Sec. \ref{conc}

\section{Theory and Approach}
\label{par:2}
\subsection{Modified VFF method}
The Valence Force Field (VFF) model is a force constant based atomic potential (U) calculation method. The MVFF model we present here is based on the combination of two different extended VFF models, (a) VFF model from Sui et. al. \cite{VFF_mod_herman} which is suited for non-polar materials like Si, Ge and (b) VFF model from Zunger et. al. \cite{VFF_mod_zunger} which is suited for polar materials like GaP, GaAs, etc. These methods have successfully modelled the phonon dispersion in various semiconductor materials \cite{Keating_VFF,VFF_mod_herman,VFF_mod_zunger}. The force constants represent the various kinds of interaction between the atoms. For phonon modeling in diamond (and zinc-blende) lattices we consider the following interactions (Fig.~\ref{fig:VFF_short_range}), (a) bond-stretching ($\alpha$), (b) bond-bending ($\beta$) (c) bond stretch coupling ($\gamma$) (d) bond bending-stretching coupling ($\delta$) and (e) co-planar bond bending coupling ($\lambda$). The first two terms (a and b in Fig. \ref{fig:VFF_short_range}) are from the original Keating model \cite{Keating_VFF}, which are not sufficient to reproduce the bulk phonon dispersion accurately in zinc-blende semiconductors \cite{VFF_mod_herman,jce_own_paper}. Hence, higher order interactions (term c, d and e in Fig. \ref{fig:VFF_short_range}) are included in the original KVFF model. The force constants in the MVFF model depend on the bond length and angle variations from the ideal value \cite{VFF_mod_herman}. Due to this reason MVFF is also known as the `quasi-anharmonic model'.

The motion of atoms in the semiconductor structure is captured by a dynamical matrix (DM). The dynamical matrix is assembled using the second derivative (Hessian) of the crystal potential energy (U). The bulk DM has periodic boundary conditions along all the directions (Born-von Karman condition, Fig.~\ref{fig:VFF_short_range}) due to the assumed infinite material extent along all the directions. For free-standing SiNW, periodic boundary conditions are applied only along the length of the wire (X-axis) assuming an infinitely long wire, whereas the surface atoms (Y-Z axis) are free to vibrate as illustrated in Fig.~\ref{fig:Si_NW_100}). The complete calculation details are provided in Ref. \cite{jce_own_paper}.

\subsection{Lattice property calculations}
A wealth of information can be extracted from the phonon spectrum of solids. The description of the few important ones are provided below.

\textit{Sound Velocity:} One important parameter is the group velocity ($V_{grp}$) of the acoustic branches of the phonon dispersion. Near the Brillouin Zone (BZ) center $V_{grp}$ characterizes the velocity of sound ($V_{snd}$) in the solid. Depending on the acoustic phonon branch used for the calculation of $V_{grp}$, the sound velocity can be either (a) longitudinal ($V_{snd,l}$) or (b) transverse ($V_{snd,t}$).
Thus, $V_{snd}$ is given by,

\begin{equation}
\label{eq_vsnd}
	V_{snd l/t} = V_{grp} = \frac{\partial \omega(\lambda,q)}{\partial q}\Big \vert_{q\leftarrow 0},
\end{equation}
where, $\lambda$ and q are the phonon polarization (for longitudinal or transverse direction) and wave vector respectively.

\begin{figure}[t!]
	\centering
		\includegraphics[width=3.0in,height=2.1in]{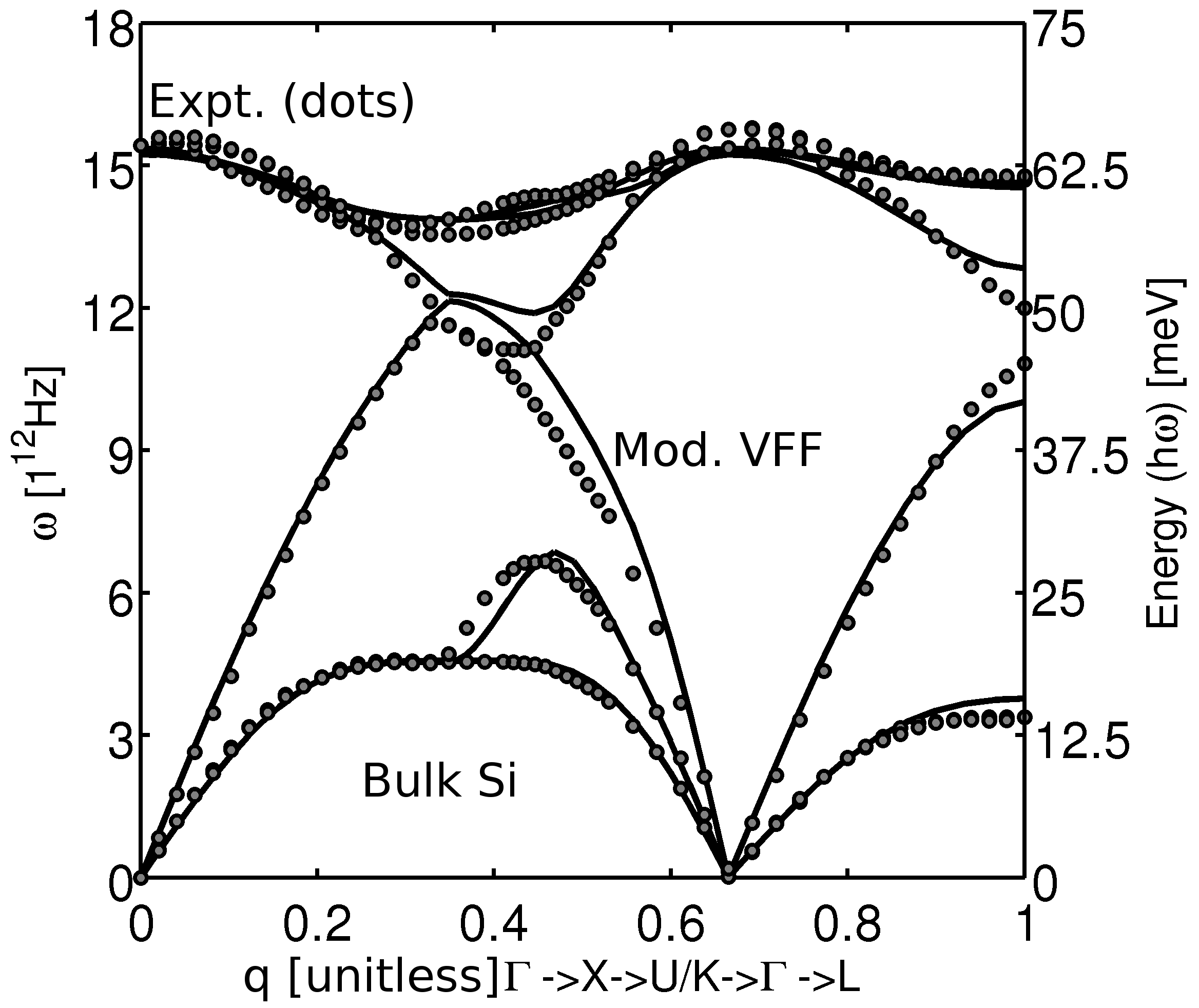}
	\caption{Comparison of the MVFF method (lines)  and the experimental data (dots) \cite{bulk_si_exp_phon} for bulk Silicon phonon dispersion at 80K.}
	\label{fig:VFF_benchmark}
\end{figure} 

\textit{Lattice thermal properties:} In a semiconductor where its two ends are maintained at a small temperature difference ($\Delta T$), its specific heat ($C_{v}$) and thermal conductance ($\kappa_{l}$) can be evaluated from the phonon dispersion using the Landauer's formula \cite{Land}. The temperature dependent specific heat is given by\cite{Fph_1,sinw_cv},

 \begin{equation}
\label{eq_sp_heat}
 	C_{v}(T) = k_{B} \sum_{n,q} \Big[ \frac{ \Big(\frac{\hbar \omega(n,q)}{k_{B}T}\Big)^{2} \cdot \exp\Big(\frac{-\hbar \omega(n,q)}{k_{B}T}\Big)} {(1-\exp\big(\frac{-\hbar \omega(n,q)}{k_{B}T)}\big)^{2}} \Big],
\end{equation}
where $k_{B}$, $\hbar$, T, and n are Boltzmann's constant, reduced Planck's constant, mean temperature, and number of sub-bands, respectively. The temperature dependent lattice thermal conductance ($\kappa_{l}$) for a 1D conductor is given by \cite{Mingo,jauho_method,Fph_1},

\begin{equation}
\label{eq_kbal}
 	\kappa_{l}(T) = \hbar \int^{\omega_{max}}_{0} M(\omega)\cdot \omega \cdot \frac{\partial}{\partial T}\Big[(\exp\big(\frac{\hbar\omega}{k_{B}T})-1\big)^{-1} \Big]d\omega,
\end{equation}
where, $M(\omega)$ is the number of modes at a given frequency $\omega$. In the next section we present the results on phonon dispersion in SiNWs and the lattice thermal properties.

\begin{table}[b!]
\caption{Force constants used for Silicon in MVFF model} 
\vspace{6pt}
\begin{center}
 \begin{tabular}{|l|c|c|c|c|r|}
 \hline
 Sp.const(N/m) & $\alpha$&$\beta$  & $\gamma$ &$\delta$  & $\lambda$ \\ \hline
 Silicon & 45.1 & 4.89& 0 & 1.36 & 9.14 \\
 \hline
\end{tabular}
\end{center}
\label{table1}
\end{table}

\section{Results and Discussion}
\label{sec:3}

\subsection{Experimental benchmark}
\label{par:3}
The MVFF model has been bench marked by calculating the phonon dispersions in different bulk zinc-blende semiconductors. One of the representative calculation is shown in Fig.~\ref{fig:VFF_benchmark} for bulk Si. The model accurately reproduces the important features of the experimental phonon dispersion of Si (at 80K) \cite{bulk_si_exp_phon} in the entire BZ. The force constant values for Si are listed in Table.\ref{table1}. The bond stretch coupling parameter ($\gamma$) is zero for Si. This term contributes significantly for splitting the longitudinal optical (LO) and longitudinal acoustic (LA) phonon branches near the BZ edge only in III-V materials like GaAs, GaP \cite{VFF_mod_zunger}.

\subsection{Nanowire phonon dispersions}
\label{par:4}

\begin{figure}[t!]
	\centering
		\includegraphics[width=3.32in,height=2.3in]{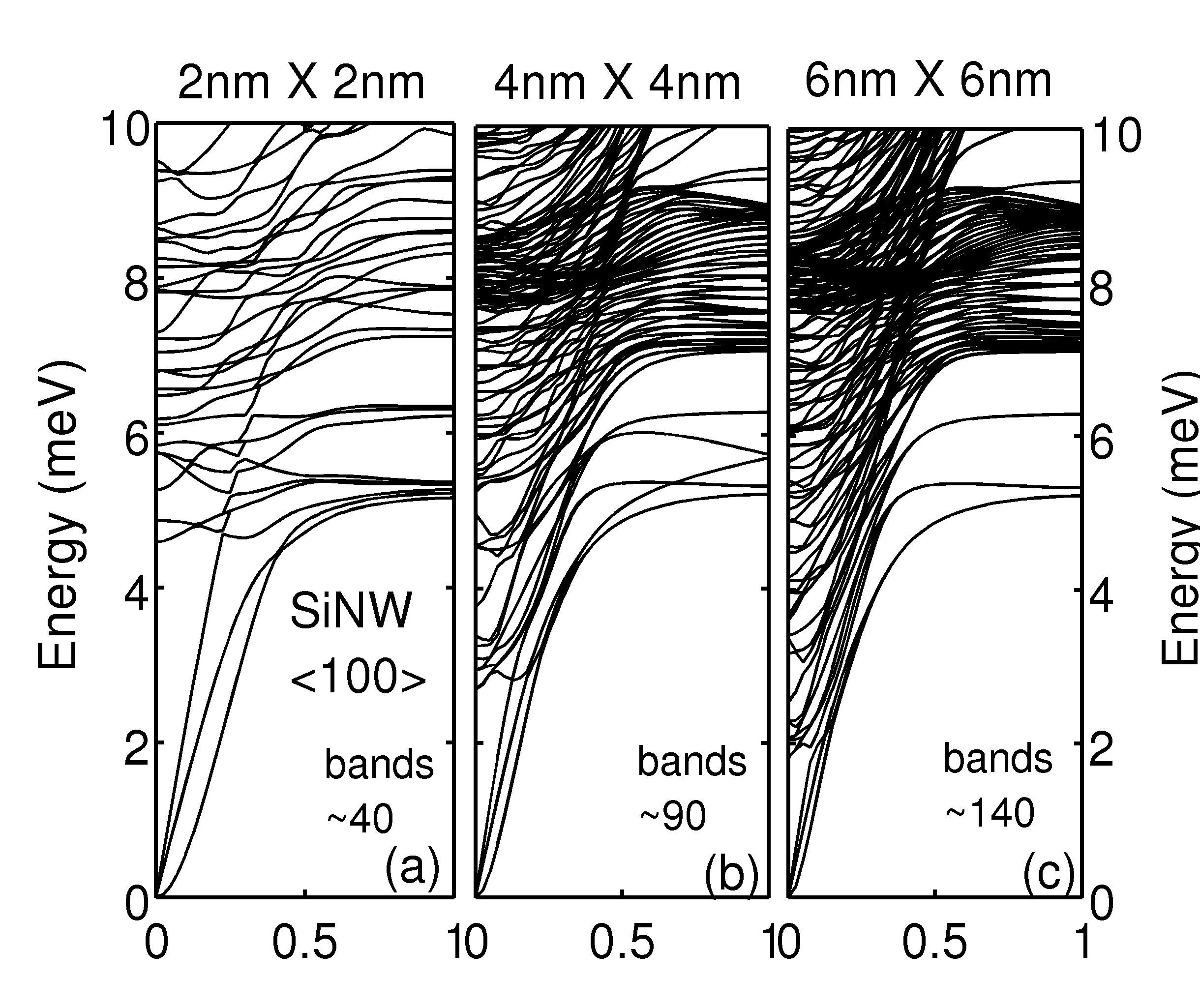}
	\caption{Effect of cross-section size on the phonon dispersion in $\langle$100$\rangle$ SiNW. 
	        The wire dimensions are (W = H) (a) 2nm (b) 4nm, and (c) 6nm. As the cross-section size increases the number of phonon modes within a given energy range increase as shown here. The geometrical quantization increases the separation of bands in smaller wires.}
	\label{fig:phonon_disp_NW}
\end{figure} 

\begin{figure}[b!]
	\centering
		\includegraphics[width=3.0in,height=3.4in]{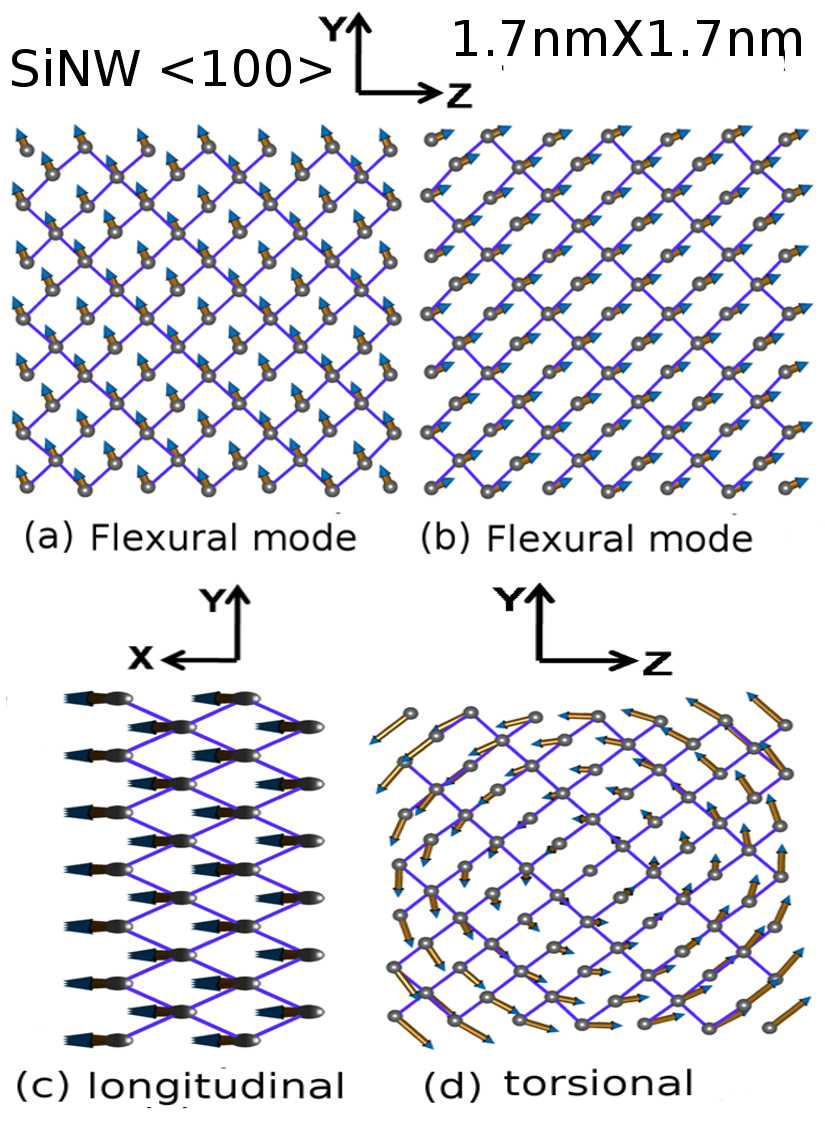}
	\caption{Eigen modes of the atomic displacement shown in a SiNW with W, H = 1.63nm. 
	 The atomic displacement are shown at $q_{x} \sim$  0 for (a) mode = 1, (b) mode = 2. These modes are called \textit{flexural modes} and cause bending of the NW. (c) mode = 3, called the \textit{longitudinal mode} causes atomic motion along the wire axis. (d) mode = 4, called the \textit{torsional mode}, observed only in wires, represents the extra rotational degree of freedom. }
	\label{fig:eigen_disp}
\end{figure} 
The phonon dispersion is calculated in free-standing $\langle$100$\rangle$ SiNW where the surface atoms are allowed to vibrate freely (Fig.~\ref{fig:Si_NW_100}). The dispersion in SiNWs with cross-section size (W=H) (a) 2nm, (b) 4nm, and (c) 6nm are compared in Fig.~\ref{fig:phonon_disp_NW}. The number of sub-bands increases as the wire cross-section size increases due to an increased number of atoms (Fig.~\ref{fig:phonon_disp_NW}). This results in a higher degree of vibrational freedom. A comparison of the number of sub-bands within 10 meV of energy for the different SiNWs considered here shows that the 2nm $\times$ 2nm structure has $\sim$ 40 sub-bands, the 4nm $\times$ 4nm has $\sim$ 90 sub-bands and the 6nm $\times$ 6nm has nearly 110 sub-bands (Fig.~\ref{fig:phonon_disp_NW}). The increased separation of phonon sub-bands in smaller wires is a consequence of the geometrical confinement. The higher sub-bands have mixed acoustic and optical branch like properties. Many of the sub-bands in this regime have group velocities close to zero which again indicates a strong phonon confinement in thin nanowires. In SiNWs these flat phonon branches appear due to the zone-folding of acoustic branches from the bulk phonon-dispersion (Fig.~\ref{fig:VFF_benchmark}).

\textit{Eigen modes of oscillation:} The first four vibrational Eigen modes are shown in Fig.~\ref{fig:eigen_disp} (arrows show the displacement direction). The first two modes (Fig.~\ref{fig:eigen_disp} a, b) are the flexural modes responsible for NW bending perpendicular to the NW axis. The Eigen frequencies for the lowest two phonon bands near the zone center are $\propto q^2$.  The next mode (Fig.~\ref{fig:eigen_disp} c) is the longitudinal mode. The mode no. 4 is torsional mode which represents the extra degree of rotational movement in free standing wires (Fig.~\ref{fig:eigen_disp} d). Mode no. 3 and 4 behave like the bulk acoustic branches ($\propto q$ near the zone center) \cite{mahan_phonon}. Mode 1 and 2 appear due to the free-standing nature of the nanowires which are not available in bulk. Hence, new modes as well as new phonon dispersion branches appear in SiNW which are very different from their bulk counterpart. This also has a strong impact on the lattice properties of the wires which is discussed in the following part.

\subsection{Lattice properties in SiNW}
\label{sec_lat_prop}
\begin{figure}[t!]
	\centering
		\includegraphics[width=3.2in,height=2.2in]{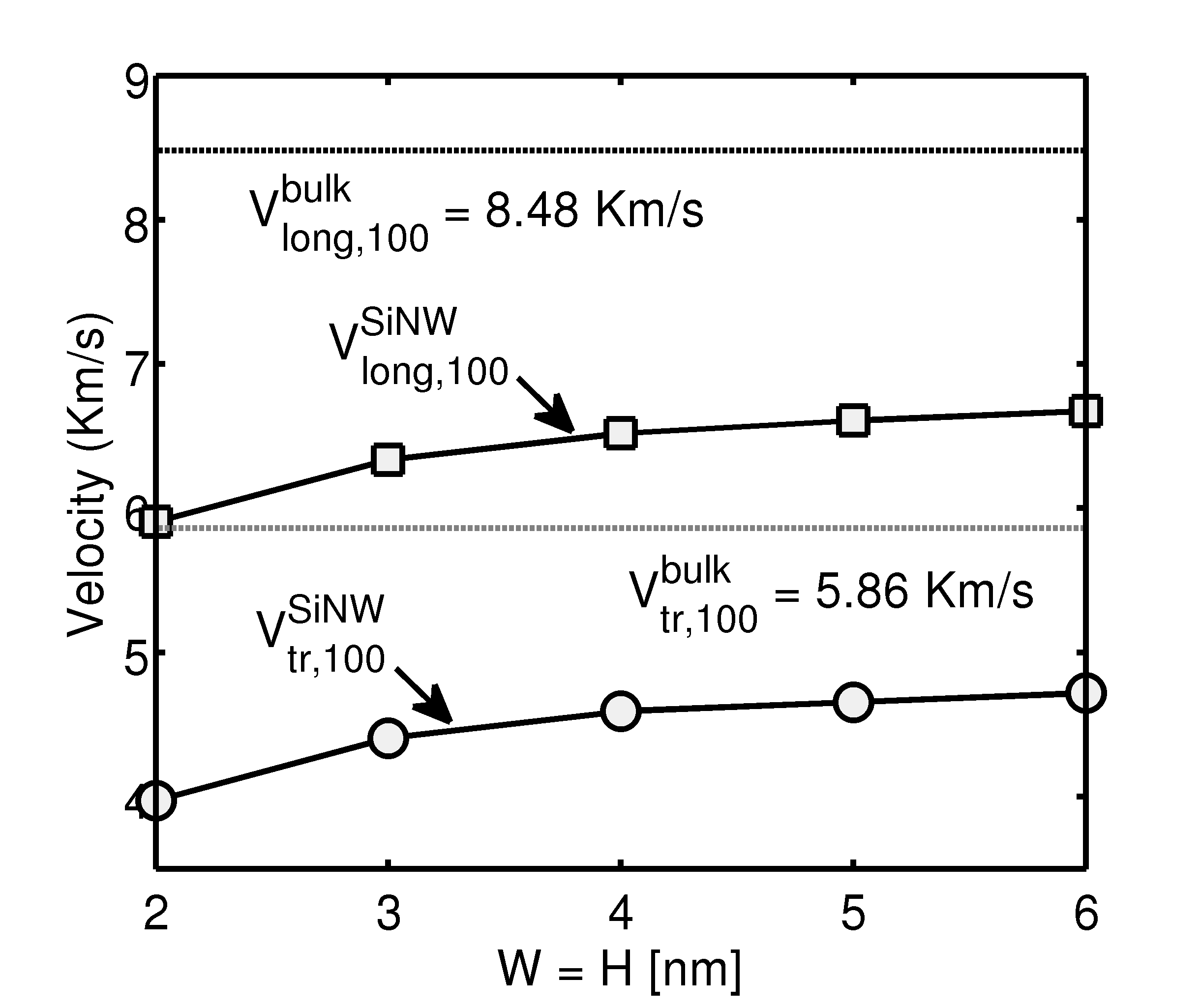}
	\caption{Sound Velocity ($V_{snd}$) in $\langle$100$\rangle$ SiNW with free boundary. As a reference the bulk $V_{snd}$ is shown along the $\langle$100$\rangle$ direction \cite{Holland}. Increasing the wire diameter increases the $V_{snd}$ since the vibrational degree of freedom increases.}
	\label{fig:vsound_dep}	
\end{figure}

\paragraph*{\textbf{\underline{Sound velocity variation ($V_{snd}$)}}} 
The sound velocity ($V_{snd}$) is another way to compare the vibrational modes in SiNWs. $V_{snd}$ is obtained using the sub-bands 3 and 4 (transverse and longitudinal modes) of the phonon dispersion.
$V_{snd}$ shows a reduction with decreasing wire cross-section size (acoustic mode softening) \cite{mahan_phonon} (Fig.~\ref{fig:vsound_dep}). This happens since the acoustic branches get flatter in small wires due to phonon confinement. This strongly affects the thermal properties of the SiNWs since the acoustic branches are mainly responsible for heat conduction. A smaller sound velocity also results in smaller thermal conductance (discussed later).   

\begin{figure}[b!]
	\centering
		\includegraphics[width=3.2in,height=2.4in]{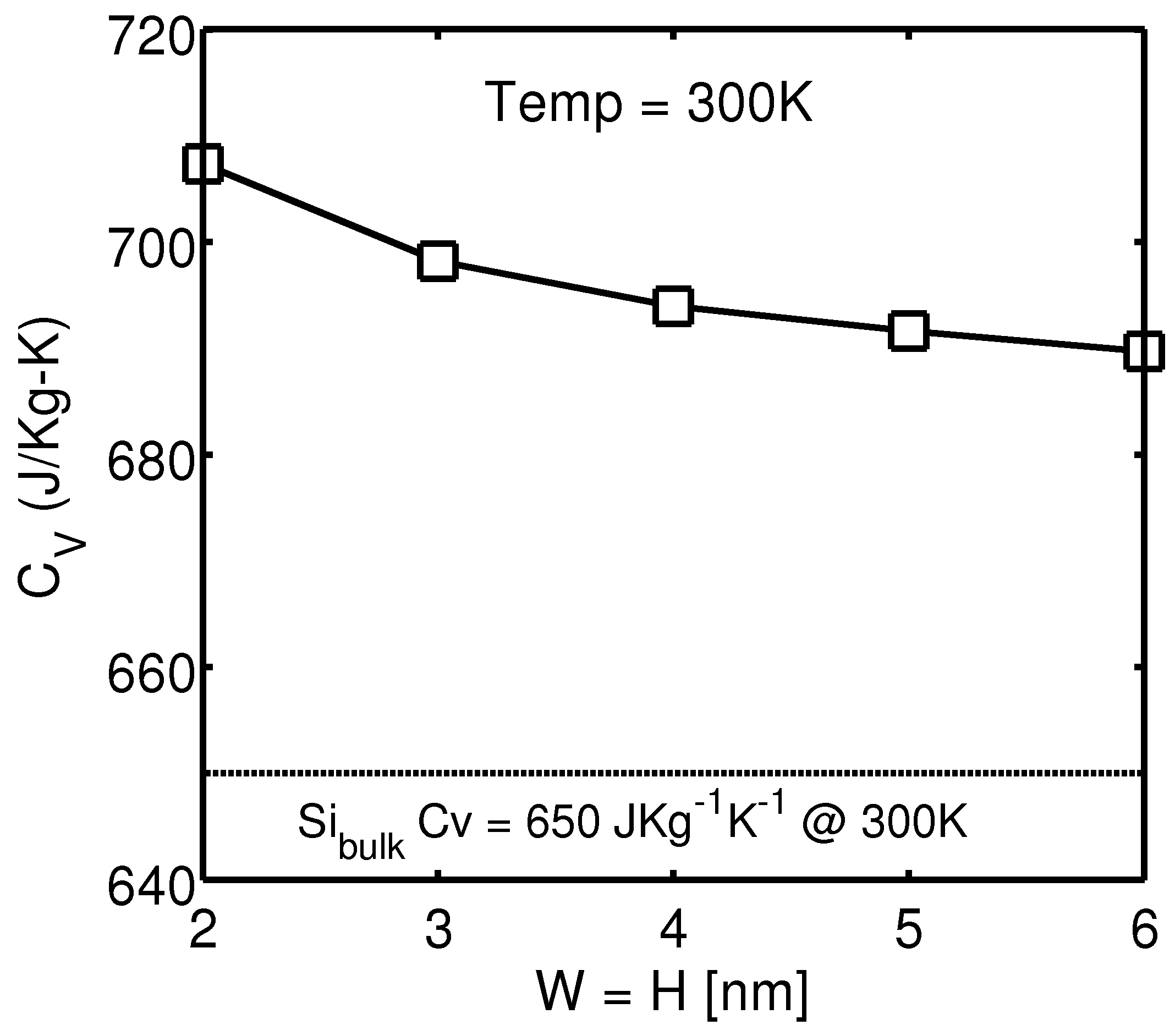}
	\caption{Variation of the specific heat ($C_{v}$) in Silicon nanowires with width. As a reference the specific heat for bulk Si is shown by a dashed line. SiNWs have larger $C_{v}$ than bulk Si \cite{sinw_cv}. The specific heat increases as the cross-section size of the wire decreases.}
	\label{fig:phonon_cv}
\end{figure}

\paragraph*{\textbf{\underline{SiNW specific heat ($C_{V}$) variation}}}
The specific heat of silicon nanowires increases with their decreasing cross-section size (Fig.~\ref{fig:phonon_cv}). This can be attributed to two reasons, (i) phonon confinement due to small cross-section and (ii) an increased surface-to-volume ratio (SVR) in smaller wires. Geometrical confinement separates the phonon bands in energy with decreasing wire cross-section size (Fig.~\ref{fig:phonon_disp_NW}) which makes only the few lower energy bands active at a given temperature (see Eq.\ref{eq_sp_heat}). This explains the increasing energy needed to raise the temperature of the smaller wires. Also the increasing SVR results in a higher partial phonon DOS associated with the wire surface, which further enhances the specific heat with decreasing wire cross-section \cite{sinw_cv}.

\begin{figure}[t!]
	\centering
		\includegraphics[width=3.2in,height=2.0in]{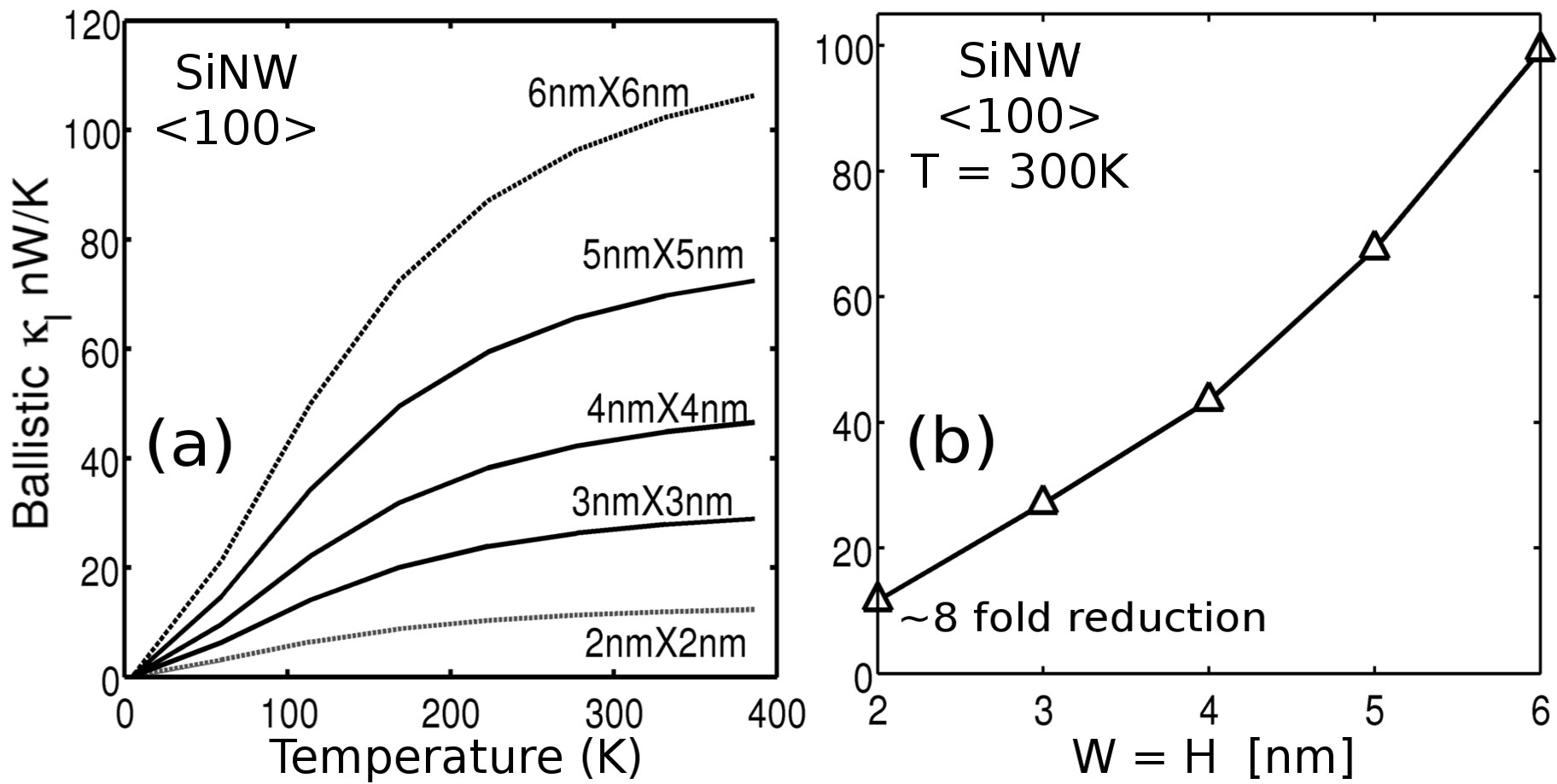}
	\caption{Variation in the ballistic lattice thermal conductance ($\kappa_{l}$) (a) with temperature for different cross-section size SiNW and (b) at room temperature (300K) for different NW width size.}
	\label{fig:kbal_dep}
\end{figure} 

\paragraph*{\textbf{\underline{Lattice thermal conductance ($\kappa_{l}$) variation}}}

The modified phonon dispersion in SiNWs significantly affects the ballistic $\kappa_{l}$ \cite{Mingo}. The ballistic $\kappa_{l}$ increases with temperature as shown in Fig. \ref{fig:kbal_dep}a. 
The spread of the Bose-Einstein distribution ($F_{BD} = [\exp(\hbar\omega/k_{B}T)-1]^{-1}$ in energy increases with increasing temperature. This results in a larger number of modes contributing to the heat transfer \cite{jauho_method} (see Eq. (\ref{eq_kbal})). This explains the increased ballistic $\kappa_{l}$ at elevated temperatures. Larger wires show more $\kappa_{l}$ since, (i) they have more phonon sub-bands resulting in more number of modes (see Eq. (\ref{eq_kbal})) and (ii) they have a higher acoustic velocity which is responsible for larger heat conduction (Fig. \ref{fig:vsound_dep}). The ballistic $\kappa_{l}$ of 2nm $\times$ 2nm Si nanowire reduces by a factor of 8 as compared to the 6nm $\times$ 6nm SiNW (Fig. \ref{fig:kbal_dep}b).

An important point to note here is that $\kappa_{l}$ is expected to decrease further in smaller wires due to phonon scattering by other phonons, interfaces and boundaries \cite{Mingo} which is neglected in the present study. The main message here is that even ballistic phonons show a reduction in thermal conductance with narrow SiNWs, which comes from (i) a modification of the phonon dispersion (ii) phonon confinement effects in SiNWs.

\section{Conclusions}
\label{conc}
A modified atomistic force constant based model has been developed for the calculation of phonon dispersion in zinc-blende lattices. This model shows a very good agreement with experimental bulk phonon data. An important impact of device miniaturization is the presence of surfaces (and hence higher surface-to-volume ratio) which affects the phonon dispersion in  SiNW. The clear demarcation of phonon bands as acoustic and optical branches become vague in SiNW. This manifests itself in the form of modified lattice properties in SiNWs compared to bulk Si. We find that decreasing wire cross-section reduces the sound velocity and the lattice thermal conductance whereas the specific heat increases. 
The geometry dependent calculation of phonon dispersion in small nanowires is therefore important to properly understand their thermal properties.

\section*{Acknowledgments}
The authors would like to acknowledge the computational resources from nanoHUB.org, an National Science Foundation (NSF) funded, NCN project. Financial support from MSD Focus Center, one of six research centers funded under the Focus Center Research Program (FCRP), a Semiconductor Research Corporation (SRC) entity, by the Nanoelectronics Research Initiative (NRI) through the Midwest Institute for Nanoelectronics Discovery (MIND) and NSF PetaApps, grant number OCI-0749140, are also acknowledged.

\bibliographystyle{IEEEtran}
\bibliography{refs}

\end{document}